\documentclass[11pt,a4paper]{article}
\usepackage{jcappub}
\usepackage{graphicx}
\usepackage{bm}% bold math
\usepackage{color}
\usepackage{url}
\usepackage{hyperref}

\newcommand{\dmm}{\textcolor{black}}
\newcommand{\dmma}{\textcolor{black}}

\newcommand{\eV}{{\rm eV}}

\newcommand{\MeV}{{\rm MeV}}
\newcommand{\GeV}{{\rm GeV}}
\newcommand{\TeV}{{\rm TeV}}
\newcommand{\PeV}{{\rm PeV}}
\newcommand{\EeV}{{\rm EeV}}

\newcommand{\km}{{\rm km}}

\newcommand{\s}{{\rm s}}

\newcommand{\nubar}{\bar{\nu}}
\newcommand{\Mpl}{M_{\rm Pl}}
\title{Violations of Lorentz invariance in the neutrino sector: an improved analysis of anomalous threshold constraints}

\author[a,b,c]{Luca Maccione}
\author[d,e]{Stefano Liberati}
\author[f]{David M.~Mattingly}

\affiliation[a]{DESY, Theory Group, Notkestra{\ss}e 85, D-22607 Hamburg, Germany}
\affiliation[b]{Arnold Sommerfeld Center, Ludwig-Maximilians-Universit\"at, Theresienstra{\ss}e 37, 80333 M\"unchen, Germany} 
\affiliation[c]{Max-Planck-Institut f\"ur Physik (Werner-Heisenberg-Institut), F\"ohringer Ring 6, 80805 M\"unchen, Germany} 
\affiliation[d]{SISSA, Via Bonomea 265, 34136, Trieste, Italy}
\affiliation[e]{INFN, Sezione di Trieste, Via Valerio, 2, 34127, Trieste, Italy}
\affiliation[f]{Department of Physics, University of New Hampshire, Durham, NH 03824, USA}

\emailAdd{luca.maccione@lmu.de}
\emailAdd{liberati@sissa.it}
\emailAdd{david.mattingly@unh.edu}

\abstract{
Recently there has been a renewed activity in the physics of violations of Lorentz invariance in the neutrino sector. Flavor dependent Lorentz violation, which generically changes the pattern of neutrino oscillations, is extremely tightly constrained by oscillation experiments. Flavor independent Lorentz violation, which does not introduce new oscillation phenomena, is much more weakly constrained with constraints coming from time of flight and anomalous threshold analyses.   We use a simplified rotationally invariant model to investigate the effects of finite baselines and energy dependent dispersion on anomalous reaction rates in long baseline experiments and show numerically that anomalous reactions do not necessarily cut off the spectrum quite as sharply as currently assumed.  We also present a revised analysis of how anomalous reactions can be used to cast constraints from the observed atmospheric high energy neutrinos and the expected cosmogenic ones.}

\begin{document}

\maketitle
\section{Introduction} 

Lorentz symmetry is one of the grounding pillars of special relativity and hence our present understanding of the physical world. As such, its validity has been checked by numerous experiments over the last century. Despite no experimental indications of any deviations, some approaches to quantum gravity suggest that due to the fundamental structure of spacetime Lorentz symmetry might not be an exact symmetry of nature but merely an approximate or emergent symmetry at low energies that is strongly broken above some energy scale.  Searches for any violation of Lorentz invariance (LIV) have therefore received more and more attention over the last few years as possible signals of physics beyond the standard model (see e.g~\cite{Mattingly:2005re,Kostelecky:2008ts,Liberati:2009pf,Liberati:2012tb,Liberati:2012th}).

Recently, the neutrino sector has come back to the forefront of LIV searches.  Primarily this is due to the OPERA collaboration's reported \cite{Adam:2011zb} and then retracted (see revised version of \cite{Adam:2011zb}) measurement of the time of arrival of non-oscillated $\nu_\mu$ (with a contamination of $\nubar_\mu$, $\nubar_e$ and $\nu_e$ that is estimated to be less than 5\%) over the path length $\sim730~\km$ from CERN to Gran Sasso with average energy of $\sim17~\GeV$ which was in open conflict with special relativity.
The $\nu_\mu$ seemed to arrive earlier than light would, by an amount $
\Delta t = 60.7 \pm 6.9\, {\rm (stat.)} \pm 7.4\, {\rm (sys.)}$ corresponding to an apparent propagation velocity of ${\Delta c}/{c} = (2.48 \pm 0.28\, {\rm (stat.)} \pm 0.30\, {\rm (sys.)})\times10^{-5}$ where $c$ is the low energy speed of light in vacuum, and $\Delta c = v_\nu - c$. 
After discovering a flaw in the initial measurement and the realization of a new one, the OPERA result (${\Delta c}/{c} = (2.7 \pm 3.1\, {\rm (stat.)} \pm 3.4\, {\rm (sys.)})\times10^{-6}$), together with the additional results of Icarus \cite{Antonello:2012be}, and the MINOS Collaboration analysis \cite{Adamson:2007zzb} now stand as constraints on neutrino velocities in the GeV range rather than a signal of beyond standard model physics.

The furor created by the OPERA's initial report has appropriately subsided for the greater physics community. For the quantum gravity community that focuses on possible experimental signatures of quantum gravity, however, technical issues were raised in how to analyze these types of accelerator based experiments properly.  In particular, the detailed physics of anomalous reactions, by which we mean reactions that are forbidden by energy-momentum conservation in a Lorentz invariant theory but allowed in a Lorentz violating one, and how they reduce the intensity of a particle beam from source to detector became much more important.  The primary theoretical objection to the initial OPERA result was produced by Cohen and Glashow~\cite{Cohen:2011hx} shortly after the OPERA report and involved just such an anomalous reaction.  Cohen and Glashow used the fact that superluminal neutrinos should emit electron-positron pairs to argue that the OPERA results were not even self-consistent: any neutrino with the speed reported by OPERA should have lost most of its energy to pair production while it propagated from CERN to the detector at Gran Sasso. The maximum energy in the beam would therefore have dropped to be below some termination energy $E_T$, and Cohen and Glashow showed that $E_T$ for the OPERA beam was less than the average $\sim17~\GeV$ energy reported by OPERA. 

The physics of the Cohen-Glashow argument is correct, however the authors did not worry about adjusting for the finite size of the baseline. A finite baseline allows for some neutrinos to undergo only one or a few pair emissions within their time of flight. Therefore the most energetic neutrinos of the injection beam can still reach the end of the baseline with an energy larger than $E_{T}$. While this was not an issue for the Cohen and Glashow result, as it was one piece of a number of experimental and theoretical concerns about OPERA~\cite{Antonello:2012hg,Stodolsky:1987vd,Longo:1987gc}, if one wishes to use accelerator time of flight experiments alone to set robust constraints on neutrino LIV, the issue must be addressed.  The infinite baseline assumption artificially limits any Lorentz violating terms to be above a certain size so that the reaction distance is much much less than the baseline.   However, this will therefore not be the tightest constraint, as constraints can still be generated even if the source particles undergo only perhaps a few anomalous reactions over the baseline.  

The primary goal of this paper is to detail the analysis and present constraints on neutrino LIV using anomalous reactions where the finite baselines have been taken into account.  As an example of the approach, we present the results of a full Monte Carlo simulation of the propagation of OPERA energy neutrinos to produce adjusted arrival neutrino spectra in the presence of anomalous reactions and a finite baseline. We do this for both the constant superluminal speed case as analyzed by Cohen and Glashow and energy dependent speeds that have been considered extensively in the literature.  We also apply another Monte Carlo simulation of neutrino propagation, but this time for ultra-high energy cosmogenic neutrinos.  Here the energies are much higher and the baselines much longer, so the corresponding constraints projected from neutrino observatories are significantly tighter.  

The reader should note that experiments that limit neutrino LIV from anomalous reactions or time of flight are almost exclusively applied to flavor independent LIV.  If neutrino LIV is flavor dependent then it changes the oscillation patterns for neutrinos.  And, as we shall describe, neutrino oscillation experiments are in general far more sensitive to LIV than either time of flight or anomalous reactions.  The reason for this is simple.  Neutrino oscillations are controlled by quantities related to the neutrino dispersion relations, namely the energy $E$ and the mass squared difference between species $\Delta m^2$, and the experiment baseline $L$ via the combination $L \Delta m^2/E= (EL) \Delta m^2/E^2$.   Anomalous reactions and time of flight are sensitive to velocities (time of flight directly and anomalous reactions via the phase space) and hence make a change when they are of order $m^2/E^2$, without any (possibly large) $EL$ amplification.  Therefore the same size LIV operator has an effect $EL$ larger for oscillation experiments than it does for anomalous reaction or time of flight constraints. Amplification, and the higher energies/longer baselines of many oscillation experiments (up to TeV range for IceCube), yield much stronger constraints of flavor dependent LIV than flavor independent LIV.  

The paper is organized as follows.  In section \ref{Framework} we introduce the effective field theory framework for Lorentz invariance violation.  As background, section \ref{Other constraints} shows in more detail how flavor dependent LIV is tightly constrained, while section \ref{TOF} describes what we can say from time of flight constraints on flavor independent LIV.  Next, in section \ref{LIV constraints} we give the flavor independent model we examine in detail, list the rates for the relevant anomalous reactions, and report the results of our Monte Carlo analysis for accelerator energies.  In section \ref{cosmogenic} we perform a similar Monte Carlo analysis for cosmogenic neutrinos, and we give our conclusions in section \ref{conclusions}. 

\section{Framework}
\label{Framework}
When dealing with departures from Lorentz invariance, several dynamical frameworks can be envisaged. Here we shall adopt an EFT approach consisting in the so called Standard Model Extension (SME) which is obtained from the Standard Model by adding all possible Lorentz breaking, gauge invariant operators.\footnote{Of course one can envisage non-EFT scenarios \cite{AmelinoCamelia:2000mn,Magueijo:2002am,AmelinoCamelia:2010pd, AmelinoCamelia:2011bm,Ellis:2009vq, Dvali:2011mn,AmelinoCamelia:2011dx,AmelinoCamelia:2012zza,Klinkhamer:2011js,Huo:2011ra} but so far or these alternative models have not reached the theoretical maturity to be suitable for casting constrains or have to be finely designed in order to avoid constraints similar to the one s we shall discuss here, see e.g.~\cite{Maccione:2010sv}. In this sense an interesting development is provided by the so called ``Relative Locality'' proposal (see e.g.~\cite{AmelinoCamelia:2011yi}).  In this paradigm, the fundamental space for classical physics is a curved momentum space rather than space-time (the latter being a derived concept). While this construction implies the abandonment of the observer-independent concept of locality it seems it could lead to theory preserving the relativity principle within a spacetime structure with an invariant minimum length.}

For what regards the neutrino sector of the SME,  the LIV operators, at any mass dimension, have been categorized in \cite{Kostelecky:2011gq}.    We are interested in a simplified subset of operators as we can then examine baseline effects in a straightforward manner.  A significant reduction in the number of terms can be achieved by requiring that the LIV operators are rotationally symmetric (the so-called ``bumblebee'' or ``fried-chicken'' model).  Since no sidereal variations have been measured for any other experiment there is no experimental reason we cannot assume that the dispersion relations for neutrinos are rotationally symmetric in some preferred frame.  We therefore for simplicity focus on the Lagrangian for neutrinos with LIV operators of mass dimension up to six involving a vector field coupled to a neutrino $\nu$ which modify the free field equations.  This allows us to examine the first CPT-odd (at mass dimension five) and CPT-even (at mass dimension 6) higher dimension operators, as opposed to the renormalizable operators that have been studied previously.  One natural choice of preferred frame is that of the cosmic microwave background, and we denote the unit vector field at rest with the CMB by $u^\alpha$.\footnote{Since the boost factor of an Earth centered frame is only $\sim10^{-3}$ \cite{Kogut:1993ag} with respect to the CMB frame, any rotation breaking effects generated by the Earth's relative motion with respect to the CMB will be naturally suppressed by a factor of $\sim10^{-3}$ relative to rotationally invariant effects and so we ignore them.  }  Instead of arbitrary rank $N$ tensors then appearing in our LIV terms, we only have powers of $u^\alpha$.  When coupling to matter, we will always de-dimensionalize by including the appropriate powers of the Planck mass $M_{\rm Pl}$ and include an arbitrary dimensionless coefficient.  Hence our constraints will be on the size of the coefficient rather than the size of $u^\alpha$. Finally, we shall also assume for simplicity that the neutrino mass eigenstates are also eigenstates of the LIV operators. With these assumptions, one has a very simple form for the dispersion relation of freely propagating neutrinos~\cite{Kostelecky:2011gq}:
\begin{equation} 
E^2 - p^2 - (m^I)^2 = \sum_{n=1}^{4} \xi^I_n \frac{|p|^n}{\Mpl^{n-2}},
\label{eq:mdr}
\end{equation}
where $\xi^I_n$ is a coefficient that depends on the terms in the underlying Lagrangian~\cite{Kostelecky:2011gq} and $I$ labels the mass eigenstate.\footnote{The dispersion relation Eq.~(\ref{eq:mdr}) has also been often considered in the literature as a test model for how quantum gravitational effects might influence infrared physics~\cite{Gambini:1998it, Carroll:2001ws, Burgess:2002tb, Barcelo:2005fc}, even without a specific Lagrangian behind it.}   The corresponding anti-particle dispersion relation is easily derived by considering the behavior of each term under CPT and given by
\begin{equation} 
E^2 - p^2 - (m^I)^2 = \sum_{n=1}^{4} (-1)^n \xi^I_n \frac{|p|^n}{\Mpl^{n-2}}.
\label{eq:antimdr}
\end{equation}
We leave the index $n$ as a free phenomenological parameter and consider the cases $n=2,3,4$ separately (the case $n=1$ would produce huge effects at low energy and is strongly constrained).  We will place constraints on each $\xi^I_n$ independently.  

One may wonder if more general dispersion relations, in which for example various terms with different $n$ are considered to contribute together, can be studied. However, this would require a strong fine tuning. Indeed, if the violation of special relativity comes in from new physics at some energy $M$, then the natural contribution of each Lorentz violating correction term changes the speed/kinematics of the neutrino by a factor $c_n\, p^{n-2}/M^{n-2}$, where $n$ is the order of the dispersion correction and $c_n$ is a dimensionless coefficient. If $M$ is high, as perhaps expected from quantum gravity, then the $c_n$'s must be very carefully chosen, i.e.~fine tuned, to contribute equally to the change in speeds/kinematics at low energies. The alternative is to have $M$, the scale of the new physics, to be at low energies such that the $c_n$ can all be of the same (perhaps small) size. This would, however, be problematic as physics at accelerator energies of 10 GeV - 10 TeV is obviously exceptionally well explored and one would still have to account for the smallness of the $c_n$.

Therefore, we assume here that there is a hierarchy of terms governed by new physics at some mass scale $M$, which we take to be $\Mpl$ and check each order correction in energy, which corresponds to the assumption that the coefficients are not fine tuned, so that naturally there is a dominant term at any given neutrino energy. Without any custodial symmetry, one would expect the relevant operator that generates $n=2$ changes in the dispersion to dominate (see e.g.~\cite{gr-qc/0403053}). We also check $n=3$ and $n=4$ as several theoretical models suggest this kind of dispersion relations to be the dominant terms. $n=3$ or $n=4$ dispersion corrections would, of course, require some other unknown physics to prevent the $n=2$ term to be dominant (see again \cite{gr-qc/0403053} and \cite{arXiv:1106.1417,arXiv:1106.6346} for a recent discussion of these issues).

As a little background for the reader, and to show that at the level of signficance acheivable by anomalous reactions the constraints on flavor (or equivalently energy eigenstate) dependent LIV are vastly tighter than constraints on flavor independent LIV, we now briefly detail the constraints from oscillation data and time of flight.

%\section{Other constraints}
\section{Oscillation constraints on flavor dependence}
\label{Other constraints}
\subsection{Formalism}
Neutrino oscillations are sensitive to differences in $E-p$ between different mass eigenstates. In standard neutrino oscillations, this difference is governed by the squared mass differences between the mass eigenstates.  With LIV \dmm{(and our assumption that the LIV eigenstates are the energy eigenstates)} oscillations are governed by the differences in the effective mass squared, denoted $(N^I)^2 =(m^I)^2+\xi^I_n p^n/\Mpl^{n-2}$.  Therefore, neutrino oscillations do not probe any absolute LIV in the neutrino sector, but rather the differences in any LIV dispersion relations between different neutrino states. 

Let us consider a neutrino produced via a particle reaction in a definite flavor eigenstate $i$ with momentum $p$.  We will treat each LIV term in $n$ separately. The amplitude for this neutrino to be in a particular mass eigenstate $I$ is represented by the matrix $U_{iI}$, where $\sum U_{jI}^\dagger U_{iI} = \delta_{ij}$. The amplitude for the neutrino to be observed in another flavor eigenstate $j$ at some distance $L$ from the source, after some time $T$  is then
\begin{equation}
A_{ij} = \sum_I U_{jI}^\dagger e^{-i(ET-pL)} U_{iI} \approx \sum_i U_{jI}^\dagger e^{-i LN_i^2/(2E)}U_{iI}\;.
\end{equation}
The transition probability can then be written as 
\begin{equation}
P_{ij} = \delta_{ij} - \sum_{I,J>I}4F_{ijIJ}\sin^2\left(\frac{\delta N_{IJ}^2L}{4E}\right)+2G_{ijIJ}\sin^2\left(\frac{\delta N_{IJ}^2L}{2E}\right)\;,
\end{equation}
with $\delta N_{IJ}^2 = N_I^2-N_J^2$ and $F_{ijIJ}$ and $G_{ijIJ}$ are functions of the mixing matrixes. In the standard formalism used by experimentalists,
\begin{equation}
\delta N_{IJ}^2 = \Delta m_{IJ}^2 + p^2\left(\frac{\Delta c}{c}\right)^{LIV}_{IJ}\;.
\end{equation}
where now 
\begin{equation}
\left(\frac{\Delta c}{c}\right)^{LIV}_{IJ} = \xi^I_n\frac{n-1}{2}\left(\frac{p}{\Mpl}\right)^{n-2}-\xi^J_n\frac{n-1}{2}\left(\frac{p}{\Mpl}\right)^{n-2}
\label{eq:dccliv}
\end{equation}

\subsection{Constraints}

There are a number of experiments over different energies and baselines that bear on LIV neutrino oscillations.  We list these below and then detail the current best constraint, which comes from IceCube.  A nice summary of neutrino oscillation observations, with particular attention to LIV, can be found in \cite{Diaz:2011ia}.  

\subsubsection{Solar neutrinos} Neutrinos produced by the Sun at $\sim$MeV energies yielded the first hint of neutrino oscillations. In the LI framework, their flux can be understood after accounting for oscillations with $\Delta m^{2}_{\odot}\simeq 7.58\times10^{-5}~\eV^{2}$ \cite{Fogli:2011qn,pdg}. In principle solar neutrinos can also be used to constrain LIV effects. However, in this case the Mikheev-Smirnov-Wolfenstein (MSW) effect~\cite{Wolfenstein:1977ue,Mikheev:1986gs} must also be taken \dmm{into account}.  %Since \dmm{other constraints prove more difficult to reconcile with the OPERA result}, we shall ignore solar neutrinos, although in principle the analysis could be done.

\subsubsection{Atmospheric neutrinos} \label{sec:oscconstraints} Muon neutrinos and antineutrinos are produced in interactions of cosmic rays with the Earth atmosphere. Experiments detect preferentially $\nu_\mu$ and $\nubar_\mu$. In this case, the survival probability is 
\begin{equation}
P_{\nu_\mu,\nu_\mu} \simeq 1 - \sin^2(2\theta_{23})\sin^2\left(\frac{\Delta m_{atm}^2L}{4E} + \left(\frac{\Delta c}{c}\right)^{LIV}_{atm}\frac{EL}{4}\right)\;,
\end{equation}
where $\theta_{ij}$ are \dmm{the} mixing angles of the neutrino mixing matrix.
Best fit values (without LIV) are: $\sin^2(\theta_{23}) = 0.42$ and $\Delta m_{atm} \simeq 2.35\times10^{-3}~\eV^2$ \cite{Fogli:2011qn,pdg}.

\subsubsection{Reactor antineutrinos} Electronic antineutrinos produced by nuclear reactors with $\sim$MeV energy \dmm{also provide relevant oscillation measurements}. The survival probability for long baseline experiments (e.g.~KamLAND \cite{Abe:2008ee}, with a baseline of about 180 km) is
\begin{equation}
P_{\nubar_e,\nubar_e} \simeq 1 - \sin^2(2\theta_{12})\sin^2\left(\frac{\Delta m_\odot^2L}{4E} + \left(\frac{\Delta c}{c}\right)^{LIV}_{\rm long}\frac{EL}{4}\right)\;.
\label{eq:reacneuosc}
\end{equation}
Best fit values obtained in the standard LI framework are: $\sin^2(\theta_{12})\simeq 0.3$ and $\Delta m_\odot^{2} = 7.58\times10^{-5}~\eV^2$ \cite{Fogli:2011qn,pdg}. Evidence of electron antineutrino disappearance was sought on much shorter baselines ($L<1$ km) as well. In this case the survival probability is ruled by a different set of parameters \cite{pdg} 
\begin{equation}
P_{\nubar_e,\nubar_e} \simeq 1 - \sin^2(2\theta_{13})\sin^2\left(\frac{\Delta m_{atm}^2L}{4E} + \left(\frac{\Delta c}{c}\right)^{LIV}_{\rm short}\frac{EL}{4}\right)\;.
\end{equation}
Observation of $\nubar_{e}$ disappearance by the Daya Bay Experiment recently yielded the measurement $\sin^{2}(2\theta_{13}) = 0.092\pm0.016 ({\rm stat.})\pm0.005 ({\rm sys.})$ in a three-neutrino framework \cite{An:2012eh}. (Later measurements by the RENO collaboration are in agreement with this determination within experimental uncertainties \cite{Ahn:2012nd}.) However, experimental facilities are not enough sensitive yet to allow LIV effects to be studied in this context.

\subsubsection{Accelerator neutrinos} At energy $\gtrsim 1~\GeV$ a few experiments with short baselines $L\simeq 1~\km$ \dmm{have provided evidence for various oscillations, including} $\nu_\mu\rightarrow\nu_e$, $\nu_\mu\rightarrow\nu_\tau$, $\nu_e\rightarrow\nu_\tau$ and their conjugates. The T2K Collaboration, with longer baseline, reported recently evidence for oscillation $\nu_\mu\rightarrow \nu_e$ \cite{Abe:2011sj} at $\sim600~\MeV$. This process is \dmm{controlled}, \dmm{as is} the short baseline reactor case, by the angle $\theta_{13}$ \cite{pdg}, hence it cannot be used to cast constraints on $(\Delta c/c)^{LIV}$.
Also MiniBooNE reported \dmm{the detection} of $\nu_e$. Interestingly, MiniBooNE finds a 3$\sigma$ excess at 300-500 MeV of $\nu_\mu\rightarrow\nu_e$ but only a 1.3$\sigma$ excess in the conjugate channel \cite{AguilarArevalo:2008rc}, hinting perhaps at CPT violation. MiniBooNE also searched for sidereal dependence of the $\nu_{e}$ signal, placing strong constraints on some combination of SME parameters \cite{Miniboone:2011yi}.

\subsubsection{Best constraint} The best constraint to date comes from survival of atmospheric muon neutrinos observed by the former IceCube detector AMANDA-II in the energy range 100 GeV to 10 TeV \cite{Kelley:2009zza}, which searched for a generic LIV in the neutrino sector~\cite{GonzalezGarcia:2004wg} and achieved $(\Delta c /c)_{IJ} \leq 2.8\times10^{-27}$ at 90\% confidence level assuming maximal mixing for some of the combinations $I,J$. Given that IceCube does not distinguish neutrinos from antineutrinos, the same constraint applies to the corresponding antiparticles. The IceCube detector is expected to improve this constraint to $(\Delta c / c)_{IJ} \leq 9\times 10^{-28}$ in the next \dmm{few} years \cite{Huelsnitz:2009zz}. Note also that \dmm{the lack} of sidereal variations in the atmospheric neutrino flux \dmm{also yields} comparable constraints on some combinations of SME parameters \cite{Abbasi:2010kx}.

%Using the 100 GeV energy, which gives the most conservative constraints, this works out to constraints of order $\xi^I_2<O(10^{-27})$, $\xi^I_3<O(10^{-10})$, and $\xi^I_4<O(10^7)$.  In comparison, recall that the limits on neutrino speed from Icarus and Minos were of the order $\Delta c/c \leq O(10^{-5})$ and lower energies, whereas time of flight constraints from SN1987A were at the level of $\Delta c/c<O(10^{-8})$ at MeV energies.  Hence oscillation experiements are orders of magnitude more sensitive to LIV than current time of flight measurements.  As we shall show later, they are also much more sensitive than most anomalous reaction constraints as well.  
\section{Time of flight constraints on flavor independent LIV}
\label{TOF}

Starting from the seminal papers \cite{AmelinoCamelia:1997gz,Alfaro:1999wd}, the possibility of constraining LIV by simple time-of-flight observations has been explored also experimentally. In particular, the high-energy photons emitted by bursted sources at cosmological distances have led to $O(10^{-1})$ constraints on models with $n=3$ (see \cite{AmelinoCamelia:2009pg} for a review). 

When focussing on neutrinos, the following comment is in order: From here on out we attribute a definite velocity to the neutrino flavor eigenstates, although they are not energy eigenstates. However, given the above constraints on the LIV differences between neutrino energy eigenstates, and given that we are considering ultra-relativistic neutrinos, for which the mass term has negligible effect over long distances, we can safely refer to the velocity of a flavor eigenstate.

While oscillation constraints are incredibly tight, there is a class of LIV that does not induce oscillations at all.  Namely, if the LIV terms are the same for every $I$, then $\delta N_{IJ}^2$ is controlled solely by the standard PMNS matrix of mixing terms.   The dispersion relation for each neutrino is therefore identical up to the mass term, which is still $I$ dependent.

%The available techniques to establish constraints in this sector parallel the techniques used in other particle sectors of the standard model (which, after all, do not have oscillation phenomena).  \dmm{For example,} photon dispersion is constrained by absence of birefringence to the level 
%$\xi_3^{\rm ph}\lesssim 10^{-14}$
%%$M\gtrsim 10^{14}\Mpl$ 
%in the case $n=3$ \cite{Laurent:2011he, Stecker:2011ps}, and a constraint of order 
%$\xi_4^{\rm ph}\lesssim 10^{-8}$
%%$M\gtrsim 10^{4}\Mpl$
%for $n=4$ is expected from ultra-high-energy $\gamma$-rays \cite{Liberati:2009pf}. 
%Electron/positron dispersion relation is constrained by observations of synchrotron radiation from the Crab Nebula to a level 
%$\xi_3^{\rm el/pos}\lesssim 10^{-5}$
%%$M\gtrsim10^5\Mpl$ 
%for $n=3$ \cite{Maccione:2007yc}, while again ultra-high-energy cosmic ray observations help constrain the $n=4$ case to 
%$\xi_4^{\rm el/pos}\lesssim 10^{-6}$
%%$M\gtrsim10^3\Mpl$
%\cite{Liberati:2009pf}. Also in the hadronic sector constraints coming from ultra-high-energy cosmic rays are extremely strong,  for protons at the level of 
%$\xi_3^{\rm prot}\lesssim 10^{-10}$
%%$M\gtrsim10^{10}\Mpl$
%for $n=3$ \cite{Jacobson:2002hd} or 
%$\xi_4^{\rm prot}\lesssim 10^{-6}$
%%$M\gtrsim 10^3$
%for $n=4$ \cite{Maccione:2009ju}. In the case $n=2$ there is no energy scale to compare to and constraints are directly on the dimensionless parameter $\Delta c/c$. They are numerically very strong, for example $\Delta c/c_{\gamma e}\lesssim 10^{-\dmma{16}}$ \cite{Mattingly:2005re,Kostelecky:2008ts,Liberati:2012tb,Liberati:2012th}.  

The dispersion relation (\ref{eq:mdr}) implies, assuming Hamiltonian dynamics, that the propagation speed of a particle is
\begin{equation}
v(p) \simeq 1-\frac{m^2}{2p^2} + \xi_n \left(\frac{n-1}{2}\right)\left(\frac{p}{\Mpl}\right)^{n-2}\;.
\label{eq:speed}
\end{equation}
In turn the time delay (or advance) upon arrival over a path length $L$, with respect to a light ray traveling at $c \equiv 1$, is
\begin{equation} 
\Delta T(p) = \frac{v(p) - c}{c}\frac{L}{c}\;.
\label{eq:tof}
\end{equation}
We can define
\begin{equation}
\frac{\Delta c}{c}^{TOF}_{(n)} = \frac{v(p)-c}{c} = -\frac{m^2}{2p^2} + \xi_n\frac{n-1}{2}\left(\frac{p}{\Mpl}\right)^{n-2}
\label{eq:dcc}
\end{equation}

%Given that the standard convention is to include all the factors $(n-1)/2$ into $\xi_n$, we do the same in the following formulae, that can then be derived from Eq.~(\ref{eq:dcc}) by replacing $M^{n-2}\rightarrow M^{n-2}/(n-1)/2$. 
For $n=2$, $\Delta c/c$ is a parameter entering directly the modified dispersion relation. Given that $\Delta c/c$ is the quantity ``directly" accessible to the experiments at a given energy, observational constraints on $\Delta c/c$ translate in constraints on the LIV parameter via the formula
\begin{equation} 
\xi_n= \frac{2}{n-1}\left(\left.\frac{\Delta c}{c}\right|^{TOF}_{\rm Obs}+ \frac{m^2}{2p^2}\right)\times\left(\frac{\Mpl}{p}\right)^{(n-2)} 
\;.
\label{eq:mtof}
\end{equation}
%
%where the sign factor $\xi$ has been absorbed in the definition of $\Delta c/c$ to be positive.  
It is clear \dmm{from Eq.~(\ref{eq:tof}) and Eq.~(\ref{eq:mtof})} that the constraint placed on $\xi_n$ by the \dmm{measurement of a time delay} depends on both the energy and \dmm{propagation distance}.

Unfortunately we have to date only a single astrophysical event for which TOF constraints can be effectively cast on $(\Delta c/c)^{TOF}$. The explosion of SN1987a was a peculiar event which allowed to detect the almost simultaneous (within a few hours) arrival of electronic antineutrinos and photons. Although only a handful of electronic antineutrinos at MeV energies was detected by the experiments KamiokaII, IMB and Baksan, it was enough to establish a constraint $(\Delta c/c)^{TOF} \lesssim 10^{-8}$ \cite{Stodolsky:1987vd} or $(\Delta c/c)^{TOF} \lesssim 2\times10^{-9}$ \cite{Longo:1987gc} by looking at the difference in arrival time between antineutrinos and optical photons over a baseline distance of $1.5\times10^5$ ly. Further analyses of the time structure of the neutrino signal, in particular using the fact that the least energetic neutrino in the signal (at 7.5 MeV) was detected within 10 s  from the most energetic one (at 30 MeV), strengthened this constraint down to $\sim10^{-10}$ \cite{Ellis:2008fc,Sakharov:2009sh}. 

The scarcity of the detected neutrinos did not allow the reconstruction of the full energy spectrum and of its time evolution. In SN models two main effects are present and lead to the final time-energy structure of the spectrum. \dmm{First}, the SN is globally cooling, hence the average neutrino energy decreases with time; on the other hand, neutrino diffusion in the SN medium depends strongly on the energy of the neutrino, which then determines its escape time. Given these uncertainties, we find constraints purely based on the difference in the arrival time with respect to photons more conservative and robust.
%Hence, adopting $\Delta c/c \lesssim 10^{-8}$, the SN constraint implies $M \gtrsim10^8\times10~\MeV \simeq 10^{-13}~\Mpl$ for $n=3$ or $M_\nu\gtrsim0.1~\TeV \simeq 10^{-17}~\Mpl$ for $n=4$.
Hence we adopt $\Delta c/c \lesssim 10^{-8}$ which is 19 orders of magnitude weaker than the corresponding oscillation constraints.
We remark that future observations of very high-energy neutrinos from bursted sources could lead to much stronger constraints \cite{Jacob:2006gn}. 

\section{Anomalous reactions}
\label{LIV constraints}
\subsection{Background}
{
One of the most well-studied consequences of LIV, within an EFT description, is how it affects thresholds in standard reactions. Depending on the relative strengths of the LIV coefficients $\xi$ of the various particles undergoing a reaction, the energy of threshold can be different from its LI value, and a wide phenomenology of new thresholds and, possibly, of new reactions is introduced \cite{Mattingly:2002ba}. Some of these facts have been used in the past to explain, for example, the puzzling evidence (disproven later on) of protons with energy beyond the GZK threshold in the cosmic radiation \cite{Kifune:1999ex,Aloisio:2000cm,AmelinoCamelia:2000zs}, or to place strong limits on LIV in QED (see e.g.~\cite{Liberati:2009pf} for a review) and in the hadronic sector \cite{Maccione:2009ju,Saveliev:2011vw}.

Let us discuss here the new LIV threshold phenomenology relevant to our case.} Since neutrinos couple to gravity and photons (via their magnetic moment and charge radius couplings), superluminal neutrinos will emit \dmm{graviton and photon} \v{C}erenkov radiation in vacuum.  In addition, high energy superluminal neutrinos will emit neutrino/antineutrino pairs via a neutral current interaction if $n>2$.  The detailed rate computation for these reactions can be found in \cite{Jacobson:2002hd,Mattingly:2009jf}, we merely summarize below.

\begin{itemize}
\item \v{C}erenkov radiation in vacuum (photon emission): $\nu\rightarrow\nu\gamma$. This possibility has been already investigated for renormalizable operators \cite{Coleman:1998ti}. Although the rate of this reaction has always been considered too small to produce significant effects even on cosmogenic neutrinos at $10^{20}~\eV$,  the effects implied by OPERA are at much lower scales than $\Mpl$. Therefore, the rate can be strongly enhanced. The energy loss rate was computed in \cite{Jacobson:2002hd} and found to be
\begin{equation}
\tau_{\nu\gamma} \simeq \xi_{n}^{-2}\left(\frac{E}{1~\PeV}\right)^{-(2n+1)}~10^{26n-86}~\s\;.
\label{eq:vc}
\end{equation}
Gravitational \v{C}erenkov radiation can in principle be considered.
However, it is \dmm{subleading with respect to photon emission} and can possibly be an effective energy loss process only for GZK neutrinos with $E\gtrsim 10^{19}~\eV$ \cite{Alexandre:2011bu}.  \dmm{Furthermore, for $n>2$ superluminal neutrinos, neutrino splitting dominates both \v{C}erenkov reactions.}

\item Neutrino splitting ($\nu\rightarrow\nu\nu\nubar$) was studied in the context of ultra-high-energy cosmogenic neutrinos \cite{Mattingly:2009jf}. While the calculation was done for $n=4$, the methodology is readily adapted to any $n>2$. In a LI scenario the energy threshold for $\nu$-splitting would be infinite. However, with LIV there is instead a finite energy above which this reaction can happen. The threshold equation for this reaction is analytically solvable when all the neutrinos involved are in the same mass eigenstate, for which the rate is maximal.  In this case the energy threshold goes as $E_{th}=(m_\nu^2 \xi_{n}^{-1}\Mpl^{n-2})^{1/n}$.  If the outgoing neutrino/anti-neutrino pair are in a different eigenstate, then the threshold will change slightly but can still be solved for numerically. 

Using the relation $\xi_{\nubar} = (-1)^{n}\xi_{\nu}$, the typical energy loss time scale for a high energy neutrino with energy well above the threshold energy scales as \footnote{As pointed out by Ward~\cite{Ward:2012fy}, a dimensionless factor of $\xi_{n}(E/\Mpl)^{n-2}$ was missing in the original rate computed in~\cite{Mattingly:2009jf}.}
\begin{equation}
\label{eq:nusplitting}
\tau_{\nu{\rm -splitting}} \simeq \frac{ 64\pi^{3}}{3G_{F}^2E^5} \xi_{n}^{-3}\left( \frac {\Mpl} {E} \right)^{3(n-2)}\;,%\frac{ m_Z^4 \cos^2 \theta_w}{g^4E^5} \left( \frac {M} {E} \right)^{3(n-2)}\;,
\end{equation}
where $G_{F}$ is the Fermi constant. For example, for the lowest $n$ for which this reaction is allowed, $n=3$, this corresponds to 
\begin{equation}
\tau \simeq 10^{38} \xi_{n}^{-3}\left(\frac{E_\nu}{10~\GeV}\right)^{-8}~\s\;.
\end{equation}
We caution the reader that there are $O(100)$ phase space factors which vary for each $n$, so that the lifetime is only approximate. However, since the scaling with $\xi_{n}$ is so strong, these phase factors are largely irrelevant as they change $\xi_{n}$ by only an O(1) factor.

\item Neutrino pair production ($\nu\rightarrow\nu e^{+}e^{-}$) has been recently proposed in \cite{Cohen:2011hx}. The calculation was focused on the case $n=2$ as $\delta$ was considered constant when computing the terminal energy for Opera, although much of the calculation and result extends to higher $n$ with only slight modifications as we show below. {In particular, the rate for neutrino pair production of unmodified electrons will scale the same way with energy as the rate for neutrino splitting far above threshold.} The threshold energy will in general depend on $n$, and can be recovered from a scaling $\delta \rightarrow \xi_n (E/\Mpl)^{n-2}$ only up to O(1) numerical factors.

The threshold equation reads (for an electron-positron pair of opposite helicity)
%\begin{widetext}
\begin{equation}
\frac{p^{n}}{M^{n-2}}\left[\xi_{n}(1-x^{n-1})-\xi_{e}(y^{n-1}+(-1)^n t^{n-1})\right]
= m_{\nu}^{2}\frac{1-x}{x} + m_{e}^{2}\left(\frac{1}{y} +\frac{1}{t}\right)~,
\label{eq:genthres}
\end{equation}
%\end{widetext}
where $x$,$y$ and $t$ are the fraction of initial momentum $p$ carried respectively by the outgoing $\nu$, by $e^{-}$, and by
$e^{+}$, $\xi_{e}$ is the LIV coefficient of the electrons, $t = 1-x-y$ and $0<x,y,t <1$.

The computation of the threshold is straightforward but the general solution is quite cumbersome. Henceforth, we shall provide here only the form of the threshold energy for $n=3$ and furthermore assume $m_\nu\approx 0$. We also assume no violation in the electron sector $\xi_e=0$. This is justified, at least for $n=2,3$, by the strength of the synchrotron constraint \cite{Maccione:2007yc} in the electron-positron sector.
%~\footnote{Can we give generalized threshold without putting to zero the LIV in the electron sector?} 

With these assumptions, the electron/positron pair takes most of the total momentum, so that $x\simeq 0$ at threshold. This implies that the form of the threshold does not depend on $n$, apart from the scaling of the term $p^{n}/M^{n-2}$.
As a result the following general scaling holds
\begin{equation}
E_{th, (n)}^{2} = \frac{4m_{e}^{2}}{\delta_{(n)}}\;,
\end{equation}
with the replacement $\delta_{(n)} = \xi_{n}(E_{th}/\Mpl)^{n-2}$.

In addition the rate of this reaction as computed in \cite{Cohen:2011hx} is general for any $n$ up to numerical factors once one performs the same substitution as before with generic energy $\delta \rightarrow \xi_n (E/\Mpl)^{n-2}$. The generic energy loss time-scale then reads
\begin{equation}
\tau_{\nu{\rm -pair}} \simeq G_{F}^{-2}E^{-5} \xi_{n}^{-3}\left( \frac {\Mpl} {E} \right)^{3(n-2)}\;,
\end{equation}
where we dropped the purely numerical factors. \dmma{As we see, the rate matches the neutrino splitting rate Eq.~\eqref{eq:nusplitting} up to numerical factors when $n>2$.  The main difference between the two reactions is that pair production is allowed when $n=2$ while neutrino splitting is kinematically forbidden for flavor blind Lorentz breaking.}

By integrating the energy loss rate from pair production over a distance  $L$ and by assuming that the typical energy loss length be much smaller than $L$ we obtain
%\begin{widetext}
\begin{equation} \label{eq:termE}
E^{-3n+1} - E_{0}^{-3n+1} = (3n-1)\xi_{n}^{3}E_{\rm ref}^{-3(n-2)}k \frac{G_{F}^{2}}{192\pi^{3}}L \equiv E_{T}^{-3n+1}\;,
\end{equation}
%\end{widetext}
where $E$ is the energy on a neutrino starting with energy $E_{0}$ after propagation over the distance $L$  and $E_{\rm ref}$ is the energy at which we normalize the parameter $\xi_{n}$. The factor $k=25/448$ was computed in \cite{Cohen:2011hx} for the case $n=2$, while for the general case it can be found in \cite{Carmona:2012tp}. 
The ``termination'' energy $E_{T}$ corresponds to the energy that a neutrino would approach after propagation over a distance $L$ if it started with $E_{0}\gg E_{T}$. We remark here that the termination energy $E_{T}$ is a mildly varying function of $n$ and of the energy scale $E_{\rm ref}$.
\end{itemize}

\subsection{Anomalous reactions with finite baselines}

Pair production was exploited, for example, in \cite{Cohen:2011hx} to show incompatibility of the Opera result with a LIV extension of the standard model at order $n=2$. Indeed, for Opera, $\xi_{2} \sim 5\times10^{-5}$ for $E_{\rm ref}\sim 10~\mathrm{to} ~30~\GeV$, yielding $E_{T}\simeq 12.5~\GeV$. Such a small value of $E_{T}$ was incompatible with the observation of neutrinos above 40 GeV in Opera. However, Eq.~\eqref{eq:termE} does not take into account the possibility that the size of the baseline be of the same order as the energy loss length of neutrinos undergoing pair production. This allows for some neutrinos to undergo only one or a few Cherenkov emissions within their time of flight. Therefore the most energetic neutrinos of the injection beam can still  reach the end of the baseline with an energy larger than $E_{T}$. It is then necessary, in order to cast a robust constraint on LIV by using long baseline experiments, to run a full MonteCarlo simulation of the propagation of neutrinos aimed at computing the neutrino spectrum on arrival in the presence of this energy loss process. 

\begin{figure*}[tbp]
\begin{center}
\includegraphics[width=0.48\textwidth]{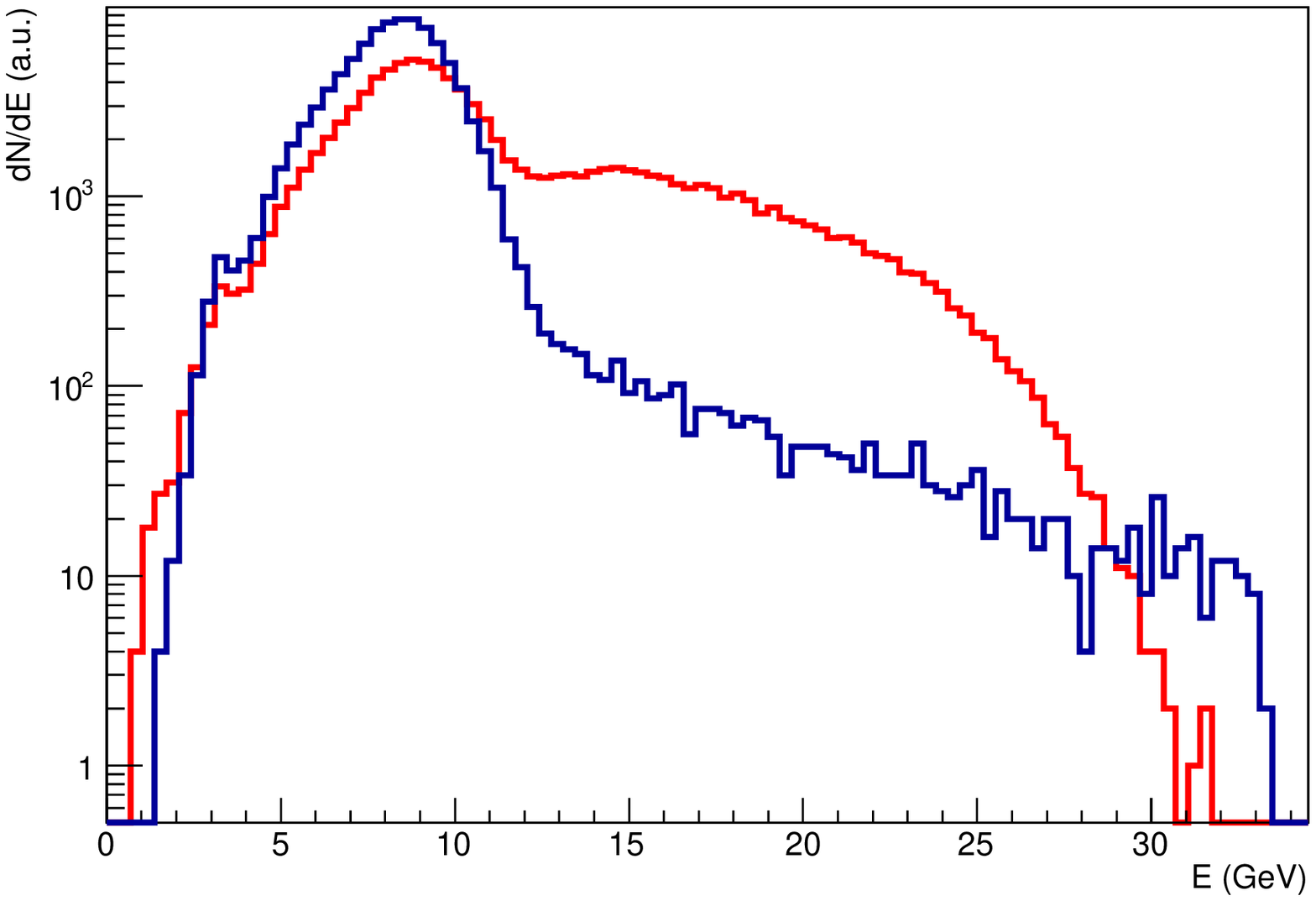}
\includegraphics[width=0.48\textwidth]{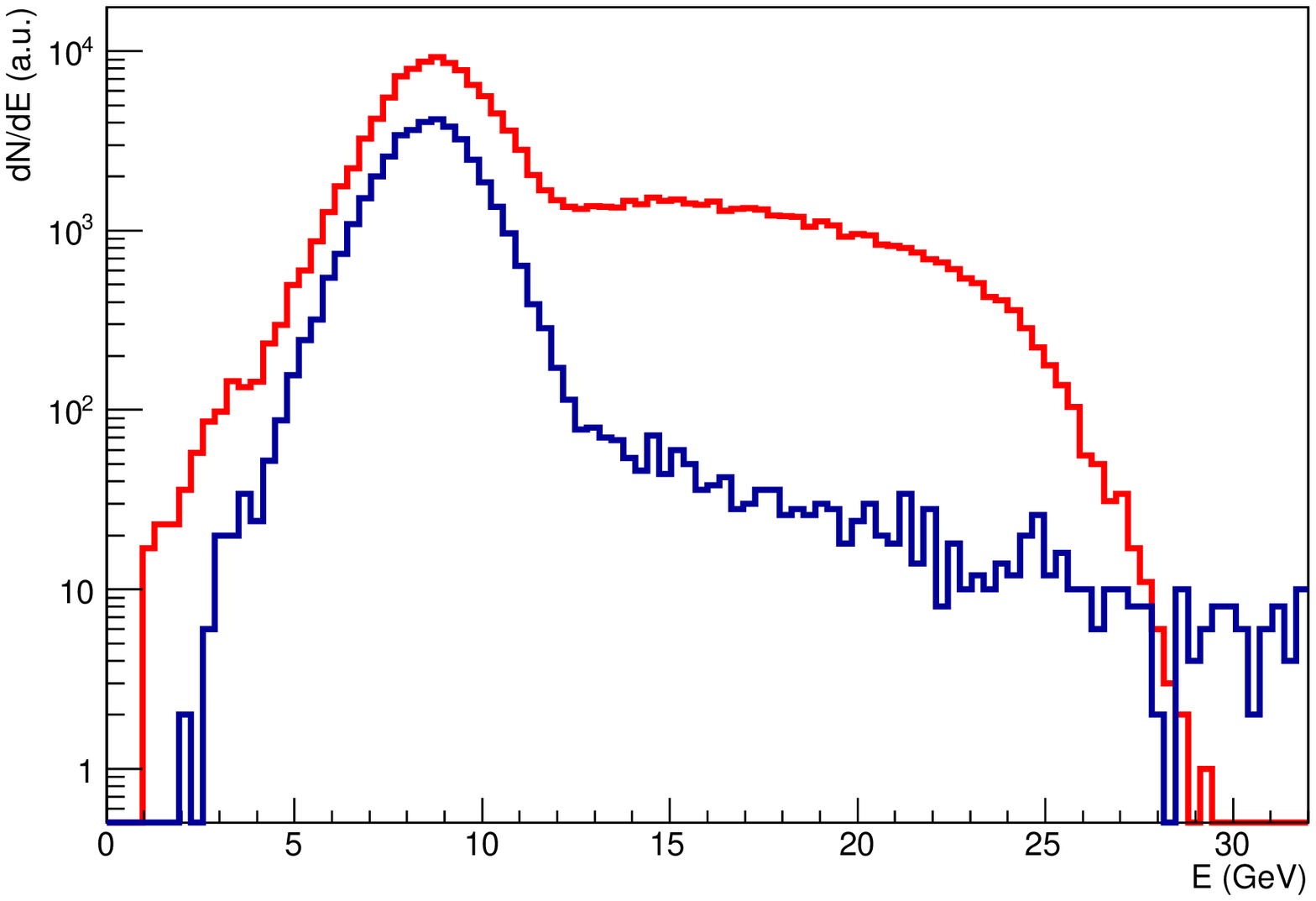}
\caption{Neutrino and pair spectra for propagation over a baseline of 730 km. In red we show the propagated neutrino spectrum, in blue the produced electron/positron spectrum. The left-hand panel refers to the case $n=2$, while the right-hand panel to the case $n=3$.}
\label{fig:completeanalysis}
\end{center}
\end{figure*}

As a paradigmatic example we show in Fig.~\ref{fig:completeanalysis} the complete analysis for the case of Opera showing the case $n=2$ and $n=3$, for which however also the process of neutrino splitting has to be taken into account. In fact, in \cite{Cohen:2011hx} the neutrino splitting process was ignored because for a flavor blind LIV in the neutrino sector this process is not kinematically allowed in the specific case $n=2$ considered there. The energy loss rate of this process is comparable to the one for pair production loss  (see Eq.~\ref{fig:completeanalysis}), and hence is not negligible for $n>2$. 

As it can be seen from Fig.~\ref{fig:completeanalysis} the propagated spectrum does indeed show a pronounced bump at the expected $E_{T}$, but is also characterized by a high energy tail that extends well above $E_{T}$ and has an amplitude about 10\% of the amplitude of the bump. 

One assumption we made in our MonteCarlo simulation was that the energy of the parent neutrino was evenly distributed among the 3 decay products. While this approximation is very well justified over most of the reaction phase space, it is not strictly true for energies close to the threshold one. However, relaxing the equipartition assumption and allowing the available energy to be not evenly distributed among the byproducts, only affects slightly the position of the peak and the shape of the spectrum, but does not sensibly change the high energy tail feature that we described above.

A separate comment is deserved for the systematics related to the uncertainties in the injection spectrum. In our reconstruction we used the injection spectrum as provided by \cite{Icarus}. We have also tested the sensitivity of the outcome to different shapes of the injection spectrum and observed that our results are little sensitive to the actual shape while keeping the average energy of the neutrino beam approximately constant at $\sim28~\GeV$.

In conclusion, this result shows that the simple calculation of $E_{T}$ is not per se conclusive for casting constraints, although our reconstruction of the propagated spectra demonstrates that in the special case of Opera the detection of neutrinos with $E>40~\GeV$ would have still pointed out an incompatibility between the adopted LIV framework and the experimental observation.

\subsection{Constraints from high energy neutrinos}
\label{cosmogenic}

The application of the aforementioned reactions is not limited to long baseline experiments such as OPERA or MINOS. Indeed, these reactions are more powerful for higher energy neutrinos and as such can be very effective in constraining LIV models beyond the energy independent $n=2$. Moreover, for $n>2$ the presence of neutrino splitting allows to cast constraints on LIV in the neutrino sector without making assumptions on LIV in other sectors (e.g.~on electron LIV).

\subsubsection{Atmospheric neutrinos}
High energy neutrinos are observed on Earth as atmospheric neutrinos up to $\sim400~\TeV$, but they are also expected as the result of photo-hadronic interactions of cosmic rays at ultra-high energy $(E\gtrsim 10^{19}~\eV)$ during their extragalactic propagation (cosmogenic neutrinos).

Constraints can be cast by considering that atmospheric neutrinos at 400 TeV need to propagate a path of the order of the Earth radius in order to be detected. By imposing that the neutrino splitting energy-loss length be larger than 6700 km at 400 TeV we would obtain a constraint of order $\xi_{3} \lesssim 40$. We note in passing that this constraint implies $\Delta c / c \lesssim 4\times10^{-17}$ at 10 GeV, much below the current sensitivity of long baseline experiments.

\subsubsection{Sensitivity estimates for UHE neutrinos}
The analysis of cosmogenic neutrino constraints requires a detailed simulation of the propagation of cosmic rays and of the products of their interactions in the intergalactic medium. Such an analysis was performed in \cite{Mattingly:2009jf}, but, as we discussed previously, the energy-loss rate was there underestimated by a factor $(E/M)^{n-2}$. 

We leave for a subsequent work the discussion of possible constraints from cosmogenic neutrinos by considering, for the flavor blind scenario of \cite{Mattingly:2009jf} --- well justified by the strong constraints on flavor dependence of LIV placed by atmospheric neutrino oscillations --- the case of pure neutrino splitting and the case of neutrino splitting and pair production. (In this second case we anticipate that the pairs would initiate an electromagnetic cascade in the intergalactic medium, leading to their energy being deposited in the GeV--TeV energy range.)

We instead correct here the order of magnitude for the expected constraint in case of detection of some UHE neutrino. This can be set by noting that these neutrinos would be produced at least 1 Mpc away from the Earth. Therefore, by using the correct rate, and imposing that the decay length for UHE neutrinos be larger than 1 Mpc, we find
\begin{equation}
\xi_{4} \lesssim 2.8\times10^{-6}\left(\frac{E}{6~\EeV}\right)^{-11/3}\;.
\label{eq:naiveconstraint}
\end{equation}

\begin{figure*}[tbp]
\begin{center}
\includegraphics[scale=0.7]{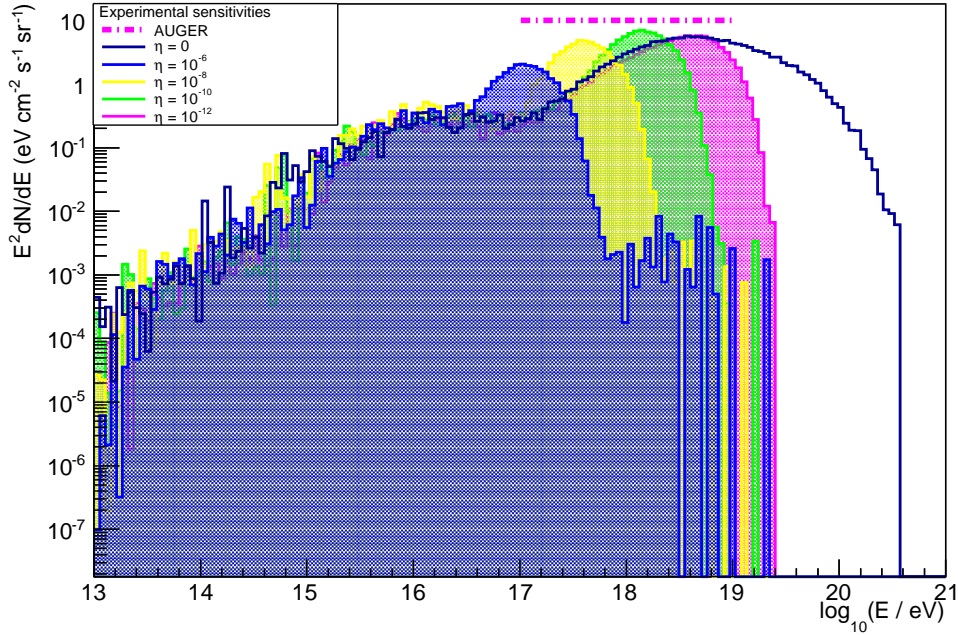}
\caption{Cosmogenic neutrino spectra and upper limits from Auger \cite{Abraham:2009uy}. The neutrino spectrum for $\xi_4=10^{-6}$ displays fluctuations at the highest energies due to poor statistics in the MonteCarlo simulation.}
\label{fig:neucosmo}
\end{center}
\end{figure*}

On the basis of our previous findings \cite{Mattingly:2009jf} we expect however that the whole UHE neutrino spectrum would be affected by LIV, as the decayed UHE neutrinos would be accumulating at around the energy threshold for neutrino splitting, thereby producing a possibly observable bump in the UHE neutrino spectrum. We evaluate this effect by running a MonteCarlo simulation of the propagation of UHE cosmic rays and their secondary products with the new framework CRPropa 2.0 \cite{Kampert:2012fi}. This framework allows to propagated UHE cosmic ray protons and nuclei in the intergalactic medium, as well as the secondary products of their interactions with the intergalactic radiation fields. 
The resulting neutrino spectra for a pure proton composition are shown in Fig.~\ref{fig:neucosmo}, together with current upper limits on the neutrino fluxes.
It is clear from Fig.~\ref{fig:neucosmo} that the LIV features in the UHE neutrino spectrum are consistent with the naive expectations from Eq.~(\ref{eq:naiveconstraint}). A bump in the UHE spectrum at sub-EeV energies is expected and could already significantly constrain $\xi_{4}< 10^{-7}$ if a measurement of UHE neutrinos could be achieved.  A simple comparison with the corresponding  constraint on  $\xi_{4}$ obtained via time of flight techniques ($\xi_{4}< 10^{34}$) should give to the reader an appropriate idea of the relative strength of the two methods although it cannot be ignored that thresholds analyses rely heavily on the specific dynamical framework differently from the more model independent TOF analysis. 

However, experimental observations of the depth of the shower maximum of UHECR interactions in the atmosphere hinted at the possible presence of nuclei heavier than protons in UHECRs \cite{Abraham:2010yv}. Given that pion production would be suppressed at UHE if heavy nuclei are a substantial component of UHECRs, the UHE neutrino flux can be much smaller than the expectation from pure proton composition. If this would be the case, a new generation of large area experimental devices would be needed in order to probe the UHE neutrino spectrum and possible LIV features.

\section{Conclusions}
\label{conclusions}

We hope that the analysis reported in this manuscript will help the reader assessing the richness, complexity, and subtlety of the possible tests of Lorentz invariance in the neutrino sector, even in this simple model which assumed rotational invariance and flavor independence (however, the latter only after considering in detail the constraints from neutrino oscillations induced by flavor dependent LIV). 

While flavor dependent LIV is strongly constrained to $(\Delta c/c)_{I,J}\lesssim O(10^{-27})$ by atmospheric neutrino oscillations, the situation is much more open for flavor independent LIV. Atmospheric neutrino observations place a constraint of order $\xi_{3}<40$, which is several orders of magnitude better than what can be achieved with time-of-flight techniques in long baseline experiments so far. In order to cast constraints of at least $O(1)$ on the case $n=4$ we need to resort to UHE cosmogenic neutrinos. The sensitivity of current experiments allows us to expect a constraint of order $\xi_{4}<10^{-7}$ if cosmogenic neutrinos will be detected.

With regards to the conclusions that can be drawn by the present study, we think that three lessons are most obvious. Firstly, in spite of the weakness of the neutrino interactions it is indeed possible to cast robust constraints using the wealth of experiments and observations dedicated to neutrinos. We can see that flavour dependent LIV is incredibly tightly constrained and that violations at the order claimed initially by OPERA are as well very difficult to accommodate in any EFT framework without unnatural fine tuning. 

Secondly, we have shown an important point to take into account when casting constraints using long baseline neutrino experiments: the finite length of the baseline does matter and one should use some moderate caution when applying reasoning \`a la \cite{Cohen:2011hx}. A full spectrum reconstruction is needed in order to cast a robust constraint. Furthermore, when extending the analysis of  \cite{Cohen:2011hx} to higher order LIV ($n>2$) one should also take into account the competing neutrino splitting mechanism (which does not give rise to electron/positron pairs as those searched for e.g. in ICARUS analysis~\cite{Antonello:2012hg} of the OPERA beam) and the energy dependence of the relevant quantities. 

Third and last, we have presented a new analysis of the constraints that can be derived from present observations of atmospheric high energy neutrinos ($E\approx 400$ TeV) and from the future detection of cosmogenic ones. While the higher energies imply stronger constraints the complexity of the analysis and the uncertainties about this fluxes require a detailed study for the reconstruction of the observed spectrum which beyond the scope of this work. We hope that the investigations reported here will be the basis for further developments in this direction.

Finally, we note that other regions of parameter space remain largely unexplored. For example we assumed from the start to perform analyses that are relatively insensitive to whether there is a Dirac or Majorana mass term. This is commonly done in LIV EFT given that the presence of CPT violation in dimension 5 operators (n=3 dispersion relation) implies different LIV terms for particles and antiparticles if isotropy/parity is preserved and hence the impossibility to identify the neutrino with its corresponding anti-particle~\cite{Kostelecky:2011gq}. One could, of course, enforce this requirement, forbid certain LIV operators, and examine the resulting kinematics on, e.g., double-$\beta$ decay experiments.  Relaxing the rotationally invariant requirement also introduces a slew of new terms which have yet to be thoroughly explored.

\begin{acknowledgments}
We are grateful to Ted Jacobson for useful remarks and suggestions. 
LM gratefully thanks Alessandro Mirizzi for enlightening discussions on neutrino physics. SL wishes to thank Matt Visser and Marco Serone for useful remarks, references and discussions. LM acknowledges support from the State of Hamburg, through the Collaborative Research program ``Connecting Particles with the Cosmos'' within the framework of the LandesExzellenzInitiative (LEXI). DM thanks the University of New Hampshire for research support and Matt Reinhold for useful discussions.
\end{acknowledgments}

\end{document}